\documentclass[prl,twocolumn,superscriptaddress,notitlepage,nofootinbib]{revtex4-2}
\usepackage[utf8]{inputenc}
\usepackage[T1]{fontenc}
\usepackage[english]{babel}
\usepackage[dvipsnames]{xcolor}
\usepackage{lmodern,fancyhdr}
\usepackage{graphicx,float}
\usepackage{enumitem,amsmath,amssymb,amsthm,bm,array,mathdots,mathrsfs,dsfont,bbold}
\usepackage[colorlinks]{hyperref}
\hypersetup{colorlinks,linkcolor=red,citecolor=blue,urlcolor=blue,final}
\DeclareMathAlphabet\mathbfcal{OMS}{cmsy}{b}{n}

\begin{document}

\title{Free evolution of the orientation of an optically levitated anisotropic nanoparticle}
\title{Rotational levitodynamics in variable degree of polarization optical tweezer}
\title{Controlling the degree of polarization in rotational levitodynamics}
\title{Experimental demonstration of free evolution in rotational levitodynamics}
\title{Full potential control in rotational levitodynamics}
\title{Full control of the libration potential in rotational levitodynamics}

\author{J.~A.~Zielińska}
\email{jzielinska@eth.ch}
\affiliation{Photonics Laboratory, ETH Zürich, CH-8093 Zürich, Switzerland}

\author{F.~van der Laan}
\affiliation{Photonics Laboratory, ETH Zürich, CH-8093 Zürich, Switzerland}

\author{A.~Norrman}
\affiliation{Photonics Laboratory, ETH Zürich, CH-8093 Zürich, Switzerland}
\affiliation{Center for Photonics Sciences, University of Eastern Finland, P.O. Box 111, FI-80101 Joensuu, Finland}

\author{R. Reimann}
\affiliation{Photonics Laboratory, ETH Zürich, CH-8093 Zürich, Switzerland}
\affiliation{Quantum Research Centre, Technology Innovation Institute, Abu Dhabi, UAE}

\author{L.~Novotny}
\affiliation{Photonics Laboratory, ETH Zürich, CH-8093 Zürich, Switzerland}

\author{M.~Frimmer}
\affiliation{Photonics Laboratory, ETH Zürich, CH-8093 Zürich, Switzerland}

\begin{abstract}
Control of the potential energy and free evolution lie at the heart of levitodynamics as key requirements for sensing, wave function expansion, and mechanical squeezing protocols.
Here, we experimentally demonstrate full control over the optical potential governing the librational degrees of freedom of a levitated anisotropic nanoparticle. 
This control is achieved by introducing the degree of polarization  as a new tool for rotational levitodynamics.
We demonstrate the free rotation of a levitated anisotropic scatterer around its short axis and and we use the rotational degrees of freedom to probe the local  spin of a strongly focused laser beam. 
\end{abstract}

\maketitle

\paragraph{Introduction.---}Levitodynamics is the science of controlling the motion of levitated mesoscopic objects~\cite{GonzalezBallesteroScience2021}.
The field has received growing attention in the last decade as a platform for force, torque, and electric field sensing~\cite{RademacherAOT2020}. 
Next to the translational degrees of freedom, the rotational dynamics of levitated anisotropic bodies offer particularly promising opportunities. More specifically, new functionalities 
demonstrated for optically levitated rotors include controllable diffusion~\cite{BellandoPRL22}, gyroscopic stabilization~\cite{KuhnNatComm2017}, spinning with GHz rotation rates~\cite{ReimannPRL2018, AhnPRL2018, AhnNatNano2020, vanderLaanPRA2020}, and the realization of rotational ``washboard potentials'' by carefully trading off conservative and non-conservative torques in elliptically polarized fields~\cite{BellandoPRL22, KuhnOptica17}. 

A particularly enticing prospect is to harness levitated rotors as ultra-sensitive torque sensors~\cite{AhnNatNano2020}, in applications ranging from photonic torque microscopy~\cite{IrreraNano2016,SvakNat18,AritaScience2020,Liang21PhotonRes,AritaArxiv2022, HuArxiv2022}, to seismology~\cite{Guattari19, Wassermann22Seis} and space-based alignment procedures~\cite{Jin2018InOrbitPE}. Another use case are tests of quantum coherence at macroscopic scales~\cite{SticklerNJP2018, SchrinskiPRA2022}. 
With librational degrees of freedom currently on track to reach the quantum regime~\cite{TebbenjohannsPRA2022, vanderLaanPRL2021, SchaeferPRL2021, PontinarXiv2022}, control over the depth and inversion of the potential will enable the generation of large delocalized orientational states~\cite{WeissPRL2021,bangbang} and the preparation of mechanical squeezed states~\cite{JanszkyPRA1992}.
 
Therefore, to realize the full promise of levitated rotors, a scheme is required to release a librator from the optical potential pinning its orientation, allowing it to freely evolve. In this state, the system becomes an optically suspended gyroscope that is extremely sensitive to DC torques, in full analogy to previously developed DC force sensing schemes~\cite{HebestreitPRL2017}. The open question is  how to deactivate the optical potential used to trap the levitated object's orientation while keeping the trapping potential for its center-of-mass (COM) motion fully intact. 

In this Letter we demonstrate full control over the conservative libration potential of an optically levitated particle. 
Our scheme makes use of the degree of polarization of the trapping field.
We experimentally realize near-zero libration frequencies up to the point where the libration signal vanishes, giving way to thermally driven free evolution of the levitated rotor. 
Additionally, for particles with cylindrical symmetry (dumbbells), we observe the signature of the transverse spin of light locally present in a strongly focused trapping beam.

\begin{figure}[b]
\includegraphics[width=0.5\textwidth]{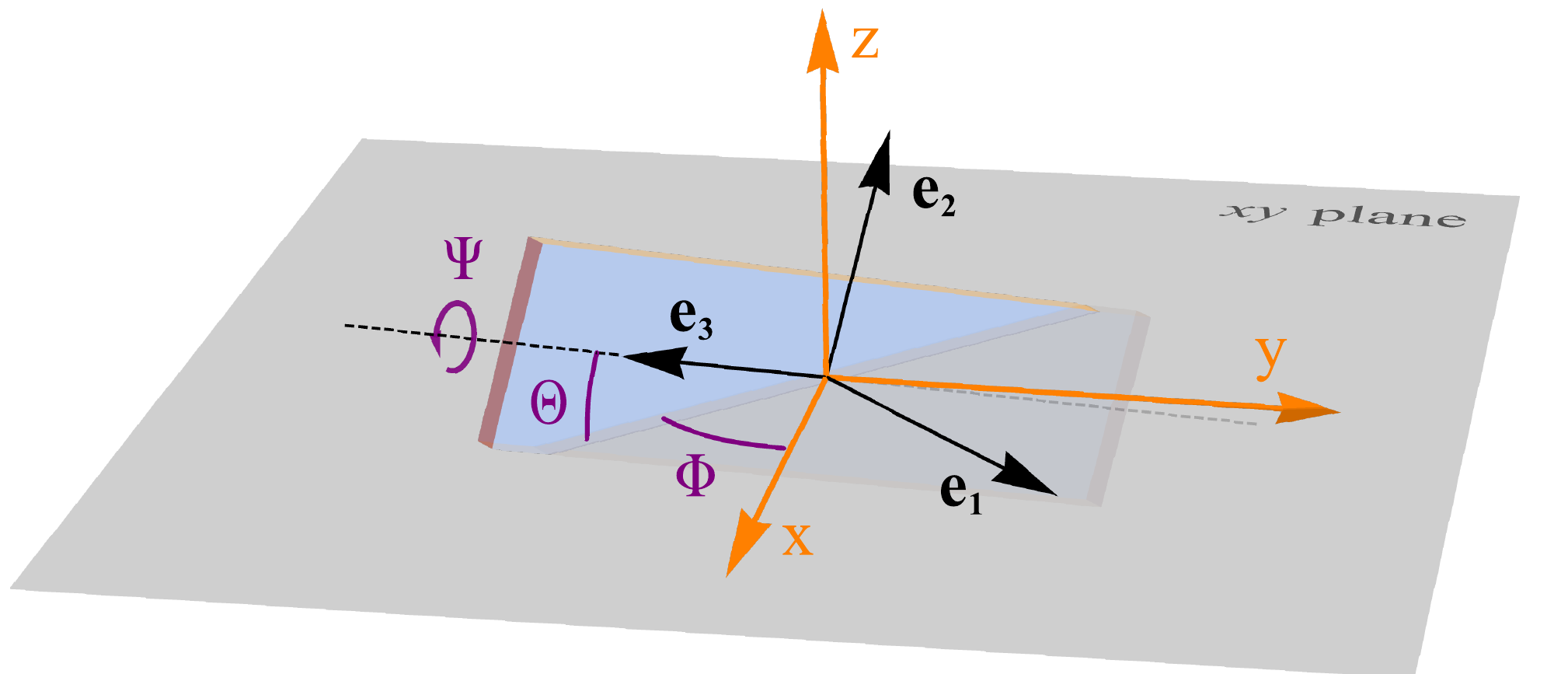}
\caption{
The orientation of an anisotropic particle's body frame (given by $\mathbf{e}_1,\mathbf{e}_2,\mathbf{e}_3$) relative to the lab frame ($x,y,z$) is described by the three angles $\Phi$, $\Theta$, and $\Psi$. 
}
\label{fig:concept}
\end{figure}

\paragraph{Key concept.---} 

Consider an anisotropic dipolar point scatterer of polarizability $\boldsymbol{\alpha}=\text{diag}(\alpha_1, \alpha_2, \alpha_3)$ in the body frame (spanned by unit vectors $\mathbf{e}_1, \mathbf{e}_2, \mathbf{e}_3$), with $\alpha_3>\alpha_2>\alpha_1$, as illustrated in Fig.~\ref{fig:concept}. 
The orientation of the particle with respect to the lab frame is described by the three Euler angles $\Phi$, $\Theta$, and $\Psi$ (see Supplement~\cite{supplementary}). 
In a field linearly polarized along $x$ in the lab frame, the particle will align with its axis of largest polarizability $\mathbf{e}_3$ to the polarization axis $x$ ($\Phi=\Theta=0$) while it can freely rotate by any angle $\Psi$ around its long axis $\mathbf{e}_3$. 
Small deviations of the long axis $\mathbf{e}_3$ from the polarization axis  represent libration modes, i.e., harmonic oscillator degrees of freedom, described by the angles $\Phi$ and $\Theta$. 

Let us now consider an unpolarized electric field, whose field vector remains in the $xy$-plane. Here, the particle will ``lie flat'' in the polarization plane, i.e., align with its axis of smallest polarizability $\mathbf{e}_1$ along the $z$ axis. Deviations from this alignment, i.e., tilts out of the polarization plane, again represent two libration modes described by the angles $\Theta$ and $\Psi$. 
At the same time, the particle can freely rotate by any angle $\Phi$, as the field vector has no preferred direction in the $xy$-plane.
Accordingly, both in a linearly and in an unpolarized field, one angular degree of freedom is free. Importantly, in the unpolarized case, the free rotation is measurable by available detection schemes~\cite{vanderLaanPRL2021, TebbenjohannsPRA2022} and therefore highly attractive for torque sensing applications. In the following, we experimentally investigate the dynamics of a levitated rotor as it is transitioned from a linearly polarized to an unpolarized trapping field.

\begin{figure}[b]
\includegraphics[width=0.48\textwidth]{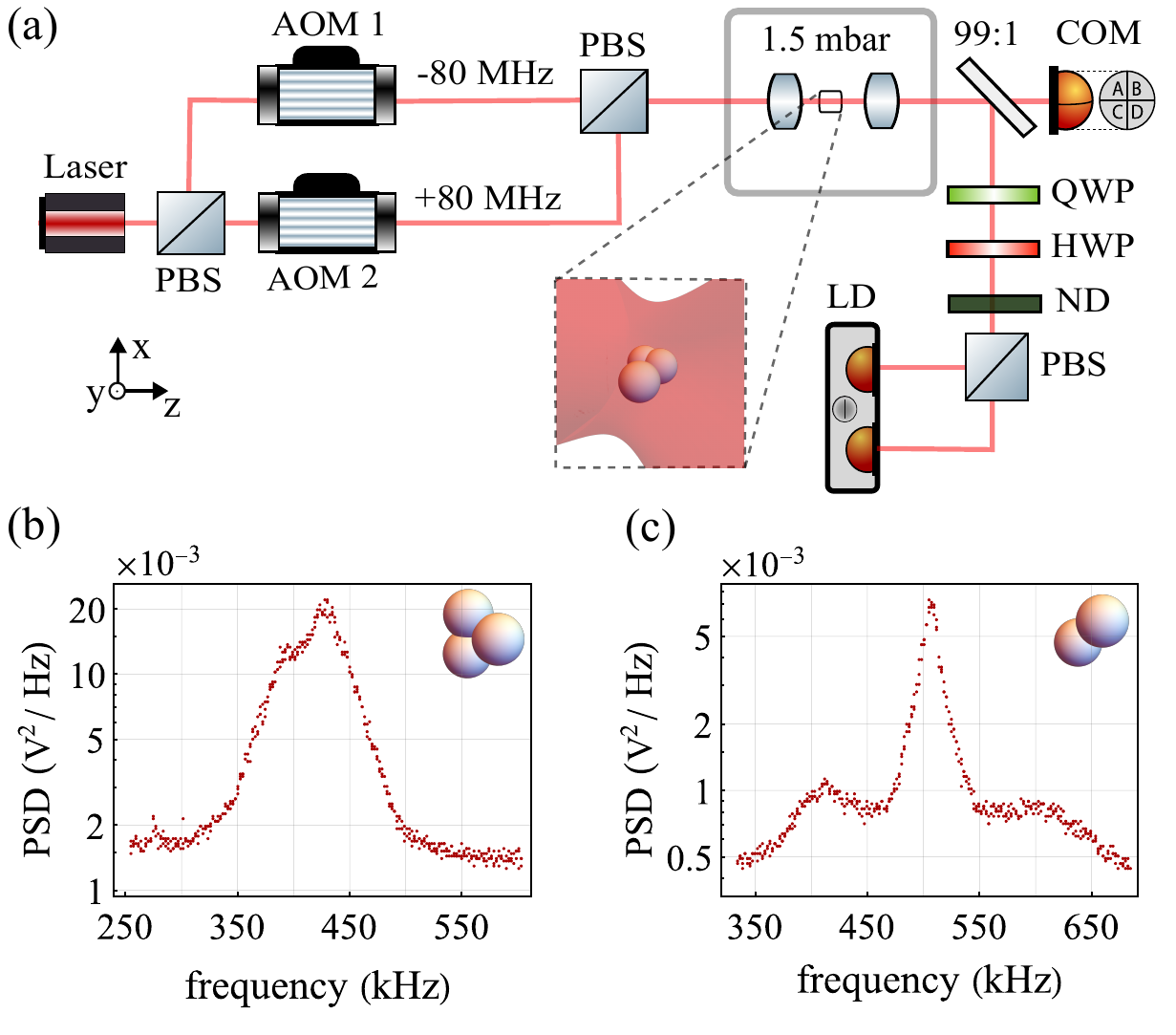}
\caption{
(a)~Simplified schematic of the experimental setup. The two polarization components of a laser beam are separated and each frequency-shifted by $\pm$80~MHz, respectively, with an acousto-optic modulator (AOM). The components' amplitudes $E_x$ and $E_y$ are varied by adjusting the driving powers of the AOMs. After recombining the polarization components on a polarizing beamsplitter (PBS), the beam is focused in a vacuum chamber with a high-NA lens to form an optical trap with variable degree of polarization. The libration signal is detected in the forward direction using a combination of a quarter-wave plate (QWP), half-wave plate (HWP), a neutral density filter (ND), a PBS, and a balanced detector (LD).
(b)~Power spectral density (PSD) of a particle cluster in a linearly polarized trap. 
(c)~PSD of a dumbbell in a linearly polarized trap. 
}
\label{fig:Setup}
\end{figure}

\paragraph{Experiment.---} At the heart of our experimental setup, illustrated in Fig.~\ref{fig:Setup}(a), is an optical trap  with variable degree of polarization (DOP), formed by focusing a trapping beam with a lens (NA=0.8) inside a vacuum chamber (pressure 1.5~mbar). To vary the DOP, a laser beam (wavelength 1550~nm) is split into two components with orthogonal linear polarization, which are then frequency shifted with acousto-optic modulators (AOMs) by $\pm80$~MHz, respectively. 
The frequency-shifted polarization components are subsequently recombined on a polarizing beamsplitter to form the trapping beam (power 450~mW), which propagates along the $z$ direction. 
Spherical silica nanoparticles (nominal diameter 143~nm) are loaded into the trap with a nebulizer. The dynamics of the trapped object are detected using forward-scattered light. The COM motion is recorded using a quadrant photodiode, and the libration signal using a standard homodyne detection scheme~\cite{vanderLaanPRL2021}.

In this work, we focus on two distinct classes of anisotropic particles, identified by their characteristic libration spectra shown  in Figs.~\ref{fig:Setup}(b) and~(c). 
The first class are ``clusters'', that is, objects composed of more than two particles, described by a fully anisotropic polarizability tensor (as the particle symbolically depicted in Fig.~\ref{fig:concept}). 
The cluster spectrum shown in Fig.~\ref{fig:Setup}(b) exhibits two modes at 390~kHz and at 425~kHz, respectively, which we associate with the libration modes described by the angles $\Theta$ and $\Phi$ from Fig.~\ref{fig:concept}. 
The second class of anisotropic particles are dumbbells (cylindrically symmetric objects composed of two spherical particles in touching contact), characterized by a sharp libration peak flanked by broad shoulders, as shown in Fig.~\ref{fig:Setup}(c). This spectrum originates from two libration modes that are coupled by the thermally driven spinning  around the symmetry axis~\cite{SebersonPRA2019, BangPRR2020, vanderLaanPRL2021}.

\paragraph{Degree of polarization.---}Having introduced the spectra for linearly polarized light we now turn to fields of variable DOP. In our setup, the tweezer field before the trapping lens reads $\mathbf{E}=(E_x e^{i \omega_x t},E_y e^{i \omega_y t},0)^T$, where the angular frequency difference $\Delta\omega=\omega_x-\omega_y=2\pi\times160$~MHz is kept constant, while the real-valued amplitudes $E_x$ and $E_y$ can be controlled with the AOMs (see Supplement for details \cite{supplementary}).
The instantaneous polarization state of the trapping beam (before focusing) is described via the four Stokes parameters~\cite{PolarizedLight16}
\begin{subequations}
\begin{align}
   \label{eq:S0}
    S_0&=E_x^2+E_y^2, \\
   \label{eq:S1}
    S_1&=E_x^2-E_y^2, \\
   \label{eq:S2}
    S_2 &=2E_x E_y \cos{(\Delta\omega t)}, \\
   \label{eq:S3}
    S_3&=-2E_x E_y\sin{(\Delta\omega t)}.
\end{align}
\end{subequations}
The DOP is defined as
\begin{equation}
\mathcal{P}= {\sqrt{\langle S_1\rangle^2 +\langle S_2\rangle^2+ \langle S_3\rangle^2}}/{\langle S_0\rangle},
\end{equation}
where $\langle\cdot\rangle$ denotes the time average~\cite{ShevchenkoOptica17}. Since the optical modulation frequency $\Delta\omega$ is more than two orders of magnitude larger than the librational dynamics, the cosine and the sine terms average out and the DOP simplifies to  $\mathcal{P}=|s_1|$, where $s_1=S_1/S_0$. The upper bound, $\mathcal{P}=1$, corresponds to fully linearly polarized light, while the lower bound, $\mathcal{P}=0$, denotes unpolarized light. The intermediate regime, $0<\mathcal{P}<1$, describes partial polarization. 

\begin{figure*}[t]
\includegraphics[width=0.9\textwidth]{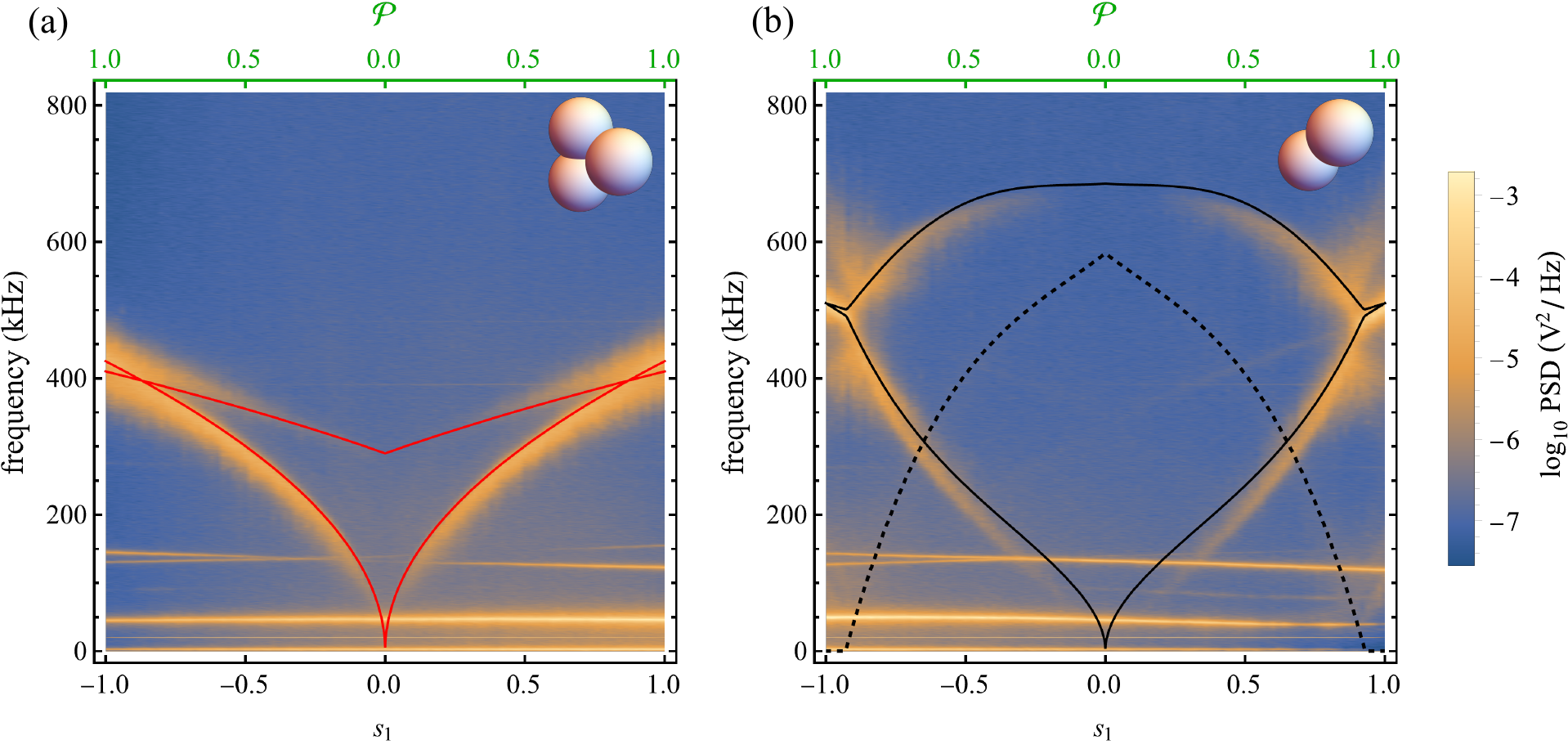}
\caption{PSDs measured by the libration detector as a function of the DOP and the normalized Stokes parameter $s_1$ of the trapping beam.  Each subfigure consists of 100 PSDs, where the first one ($s_1=-1$) corresponds to $y$-polarized trapping light and the last one ($s_1=1$) to $x$-polarized trapping light. In the case of $s_1=0$ the trapping beam is effectively unpolarized and the frequency corresponding to libration in the $xy$ plane tends to zero.  Frequencies corresponding to translational COM motion are visible as horizontal lines in the range between 40 and 150~kHz. 
(a)~Cluster (non-rotationally symmetric particle).  Red lines show theoretical prediction calculated from Eqs.~(\ref{eq:freq1}) and (\ref{eq:freq2}) using only 
libration frequencies measured at the linear polarization setting. 
(b)~Dumbbell (cylindrically symmetric particle). Black solid lines are precession frequencies calculated from the theoretical model including spinning of the dumbbell along its long axis~\cite{supplementary}, with $\omega_{\rm s}/2 \pi$ (proportional to the spinning rate) shown as dashed black line.}
\label{fig:ramps}
\end{figure*}

\paragraph{Results.---}Let us discuss our experimental observations for a  cluster trapped in a beam with variable DOP. In Fig.~\ref{fig:ramps}(a), we show in false color the power spectral density (PSD) of the libration signal as a function of frequency and $\mathcal{P}$. For linearly polarized trapping light ($\mathcal{P}=1$), we observe the spectrum from Fig.~\ref{fig:Setup}(b), with a feature composed of two closely spaced peaks near 400~kHz. As the DOP is reduced ($\mathcal{P}<1$), the two peaks split and their frequencies decrease. 
Remarkably, for unpolarized light ($\mathcal{P}=0$) the frequency of one of the modes vanishes, while the frequency of the second mode approaches 300~kHz.  

Our experimental observations for a trapped dumbbell, shown in Fig.~\ref{fig:ramps}(b), strikingly differ from that of a cluster. The single peak at 500~kHz (surrounded by broad shoulders) observed in linearly polarized light ($\mathcal{P}=1$), see Fig.~\ref{fig:Setup}(c), splits in two as the DOP is reduced ($\mathcal{P}<1$). In contrast to the cluster, the dumbbell exhibits one mode that shifts to higher frequencies and settles at 680~kHz for unpolarized light ($\mathcal{P}=0$), while the second mode frequency tends towards zero, where its signal strength vanishes. 

\paragraph{Model.---} To understand our observations, we model the orientational dynamics of an anisotropic dipolar scatterer in a field of variable DOP. Let the scatterer be characterized by its polarizability $\boldsymbol{\alpha}=\text{diag} (\alpha_1,\alpha_2,\alpha_3)$ and its tensor of inertia $\boldsymbol{I}= \text{diag} (I_1,I_2,I_3)$, which are both diagonal in the intrinsic body frame spanned by $\mathbf{e}_1, \mathbf{e}_2, \mathbf{e}_3$. We furthermore assume $\alpha_1 \leq \alpha_2 \leq \alpha_3$ and $I_1 \geq I_2 \geq I_3$. 
We calculate the potential energy of a fully anisotropic scatterer in a field of variable DOP as a function of the orientation angles $\Phi$, $\Theta$, and $\Psi$ and identify the global energy minimum (see Supplement~\cite{supplementary}). Small deviations of the orientation angles $\Phi$, $\Theta$, and $\Psi$ from their equilibrium values resemble, to first order, harmonic oscillator degrees of freedom, whose characteristic libration frequencies are given by  
\begin{subequations}
\label{eq:frequencies}
\begin{align}
   \label{eq:freq1}
   \Omega_1&= A_1\sqrt{\mathcal{P}}, \\
   \label{eq:freq2}
   \Omega_2&= A_2\sqrt{\frac{1+\mathcal{P}}{2}}, \\
   \label{eq:freq3}
   \Omega_3&= A_3\sqrt{\frac{1-\mathcal{P}}{2}},
\end{align}
\end{subequations}
respectively, 
where $A_i=[(|\alpha _j-\alpha _k| S_0)/2 I_i]^{1/2}$ and  $\{i,j,k\}$ are permutations of $\{1,2,3\}$. 

Equations~(\ref{eq:freq1})--(\ref{eq:freq3}) indicate that we can directly control the librational potential governing the orientation of the rotor via the DOP. 
Although our detection is only sensitive to libration in the  $xy$ plane~\cite{TebbenjohannsPRA2022}, the coupling between the different libration modes~\cite{SebersonPRA2019} is responsible for the second libration mode in the spectrum. 
To compare our theoretical prediction with our measurement, we plot the calculated values for the libration frequencies $\Omega_1$ and $\Omega_2$ from Eqs.~\eqref{eq:freq1} and \eqref{eq:freq2} as red lines in the measurement of the trapped cluster in Fig.~\ref{fig:ramps}(a). The required parameters $A_1$ and $A_2$ are defined by the libration frequencies extracted at $\mathcal{P}=1$. The theoretical lines trace the observed libration frequencies remarkably well, demonstrating that our model correctly captures the rotational dynamics and providing strong support for our initial assumption that the trapped object is a cluster without symmetry. 
We stress that in a field with zero DOP the cluster's libration frequency $\Omega_1$ vanishes. In other words, the orientation angle $\Phi$ undergoes free evolution.  

Let us turn our attention to the dynamics of the trapped dumbbells. Inspection of Eqs.~\eqref{eq:freq3} shows that for an object of cylindrical symmetry $A_3$ and therefore also $\Omega_3$ vanish. 
This observation intuitively makes sense, since such a scatterer can always freely rotate around its long axis. 
However, for dumbbells, the libration frequencies as a function of $\mathcal{P}$, experimentally observed in Fig.~\ref{fig:ramps}(b), deviate significantly from those predicted by Eqs.~\eqref{eq:freq1}--\eqref{eq:freq3}. 
As has been pointed out before~\cite{SebersonPRA2019,BangPRR2020}, the spinning of the dumbbell at a stationary rate $\dot\Psi_0$ around its axis of symmetry couples the libration modes with frequencies $\Omega_1$ and $\Omega_2$ into precession modes with frequencies $\Omega_{\rm A}$ and $\Omega_{\rm B}$ according to
\begin{align}
    (\Omega_{A}-\Omega_{B})^2= \omega_{\rm s}^2+(\Omega_1-\Omega_2)^2,
    \label{eq:precmodes}
\end{align}
with the coupling rate $\omega_s=\mu \dot\Psi_0 $ and the inertial coupling constant $\mu=(I_1-I_3)/I_1$. 
We interpret the salient libration features in our data in Fig.~\ref{fig:ramps}(b) as the precession frequencies $\Omega_A$ and $\Omega_B$ of the dumbbell, and fit their functional dependence with Eq.~\eqref{eq:precmodes}, where $\Omega_1$ and $\Omega_2$ are in turn given by Eqs.~\eqref{eq:freq1} and~\eqref{eq:freq2}.  The fit [black solid lines in Fig.~\ref{fig:ramps}(b)] describes our experimental observation very well and yields a fitted coupling rate $\omega_s$, shown as the dashed black line in Fig.~\ref{fig:ramps}(b)~\cite{supplementary}. 
We conclude that the rate of spinning around the long axis in our experiment reaches $\dot\Psi_0=2\pi~\times~200\,{\rm kHz}$ in the regime of unpolarized light $\mathcal{P}=0$, where we used the dumbbells length-to-diameter ratio of $L/D\approx1.8$~\cite{BellandoPRL22} to estimate the inertial coupling factor as $\mu\approx0.375$.
We will explain the origin of this spinning motion in the next section.

\paragraph{Discussion.---} Our results in Fig.~\ref{fig:ramps} demonstrate that the DOP of the trapping field allows us to control the libration frequencies of the optically levitated particle. In particular, we stress the fact that for a cluster in a field with vanishing DOP, the libration frequency $\Omega_1$ vanishes. In other words, the cluster becomes a free rotor regarding its orientation angle around the optical $z$ axis, while its two axes of largest moment of inertia ($I_2$ and $I_3$) are harmonically trapped in the $xy$ plane of polarization. The situation is analogous for the dumbbell, whose long axis is harmonically trapped with characteristic frequency $\Omega_B$ in the focal $xy$ plane in unpolarized light, while the orientation of the long axis  undergoes free evolution within this plane. Thus, the DOP allows us, for the first time,  to tune the angular motion of a levitated object from librations of several hundered kHz all the way to free evolution. This demonstration is the main result of this paper.
\begin{figure}[b]
\includegraphics[width=0.47\textwidth]{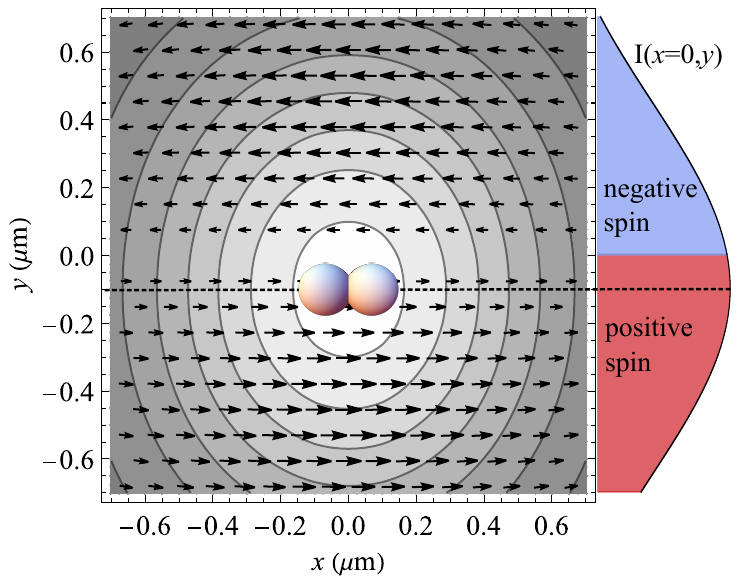}
\caption{Illustration of the spatial mismatch between the trapping beam components leading to a dumbbell being trapped in a region of nonzero transverse spin. A vector plot shows the transverse spin pattern in the $xy$ plane generated by $y$ polarized strongly focused Gaussian beam~\cite{novotny_hecht_2012}. Simultaneously, we show the dumbbell (to scale) trapped in the intensity maximum of a stronger $x$ polarized Gaussian beam (contour plot) displaced by 100 nm from $y=0$.  Beam size and lens parameters correspond to the experiment. On the right we plot the intensity profile $I(x=0,y)$, indicating regions with spin pointing in negative and positive $x$ direction. 
}
\label{fig:spin}
\end{figure}

Let us now provide an explanation for the torque that drives the trapped dumbbell into spinning motion around its long axis. 
We note that this torque must lie in the focal plane (which is the plane the dumbbell's long axis is pinned to). Strongly focused fields can indeed carry transverse spin angular momentum~\cite{EismannNat2021, ChenPRA2021}, which gives rise to a torque when transferred to a particle. 
In Fig.~\ref{fig:spin}, we illustrate the transverse part of the spin vector~\cite{GilNJP2021} in the focal plane of a strongly focused $y$ polarized Gaussian beam. The spin is depicted as arrows whose direction (length) indicates the spin's orientation (magnitude). The spin points predominantly along the positive (negative) $x$ direction in the range $y<0$ ($y>0$). 
To understand how a dumbbell can be exposed to the transverse spin, we consider a trap composed of not only a $y$ polarized beam but also of an additional strong $x$ polarized beam,  used for trapping the particle's COM (the intensity of which is illustrated as a colormap in Fig.~\ref{fig:spin}). The $x$ polarized trapping beam aligns the particle's long axis along the $x$ direction. 
If we displace  the  $x$ polarized beam along the $y$ direction then the transverse spin of the $y$ polarized beam will spin the dumbbell along its long axis, as experimentally observed. Even though the torque along the dumbbell's long axis is very weak, the effect is visible since the dumbbell is free to rotate along its long axis.
We can thus explain the observed spinning motion of the dumbbell as a signature of the transverse spin angular momentum of light in a strongly focused field, together with an inevitably imperfect alignment between the two cross-polarized beams forming our trap with tunable DOP. Effectively the trapped dumbbell locally senses the spin of an additional light field.  Note that, in contrast to  dumbbells, clusters are not driven into rotation along their long axis by a weak transverse spin of light, since $A_3\neq0$ and the angle of rotation around the long axis is restrained for $\mathcal{P}<1$ [see Eq.~\eqref{eq:freq3}]. 

Finally, let us comment on the limitations of our variable DOP potential control scheme. Throughout this work, we have only considered the mean polarization of the beams, neglecting their polarization oscillations at the frequency $\Delta\omega$ [see Eqs.~(\ref{eq:S2}) and (\ref{eq:S3})]. In analogy to Paul traps operating at RF frequencies, the oscillating polarization will give rise to an additional potential term~\cite{supplementary} and a small amplitude micromotion at $\Delta\omega$. The frequencies associated with the additional potential term are of the order of magnitude of $ B_{i}\approx A_i^2 / \Delta\omega$~\cite{GoldmanPRX2014}. In our experiment the correction to the libration frequencies caused by the oscillating polarization is negligible.  Note that $ 1/B_{i}$ limits the maximum free evolution time for a given value of $\Delta\omega$.

 Moreover, in the current implementation $\Delta\omega$ is restricted by the AOM bandwidth, but this limit can be readily removed by using two laser sources with different wavelengths (without the need for relative frequency stabilization). Increasing the free evolution time limit to 1~s requires $3~{\rm nm}$ difference between the wavelengths of the two polarization components (assuming a center wavelength of $1550~{\rm nm}$). 
 We also note that experiments involving librational potential switching~\cite{WeissPRL2021,JanszkyPRA1992} via DOP control can be implemented on timescales of a few nanoseconds (only limited by AOM rise time).

\paragraph{Conclusions.---} We have demonstrated for the first time the complete tunability of the librational frequencies of optically levitated clusters of silica nanoparticles. This tunability is accomplished by the DOP and is independent of the COM trapping potential.
Our work is important for the development of high-performance nanoscale gyroscopes and for the study of macroscopic rotational quantum physics~\cite{SticklerNature2021,SchaeferPRL2021}.
Furthermore, we have experimentally confirmed that symmetric rotors can serve as a precise tool for sensing torques not only perpendicular, but also parallel to the long axis of the rotor. This feature, together with the high control over libration degrees of freedom, may enable the full characterization of three-dimensional Stokes parameters~\cite{GilJEOS2015, IrreraNano2016, BlakemorePRA2019, GuoOptLett22}.

The authors would like to thank A. Militaru, O. Romero-Isart, C. Gonzalez-Ballestero and all trappers in the Photonics Laboratory for fruitful discussions. This research was supported by the European Union’s Horizon 2020 research and innovation programme under grant agreement No.~[863132] (IQLev), as well as ETH Grant No. ETH-47 20-2. A.N. thanks the Jane and Aatos Erkko Foundation (Finland) for funding.

\bibliography{freeevo_bibliography1}


     \pagestyle{empty}
       \setcounter{table}{0}
        \renewcommand{\thetable}{S\arabic{table}}%
        \setcounter{figure}{0}
        \renewcommand{\thefigure}{S\arabic{figure}}%
        \setcounter{equation}{0}
        \renewcommand{\theequation}{S\arabic{equation}}
        \setcounter{page}{1}
        \renewcommand{\thesection}{S\arabic{section}}
        \setcounter{section}{1}

\newpage
\pagebreak

\newcommand{\beginsupplement}{%
        \pagestyle{plain}
        \setcounter{table}{0}
        \renewcommand{\thetable}{S\arabic{table}}%
        \setcounter{figure}{0}
        \renewcommand{\thefigure}{S\arabic{figure}}%
        \setcounter{equation}{0}
        \renewcommand{\theequation}{S\arabic{equation}}
        \setcounter{page}{1}
        \renewcommand{\thesection}{S\arabic{section}}
        \setcounter{section}{0}
     }
\onecolumngrid
\newpage
\vspace{1cm}
\begin{center}
\textbf{\large Supplemental Material}
\end{center}
\beginsupplement
\section{S1. Variable DOP trapping beam preparation} 
 \label{s:calibration}
This section describes the preparation of the variable DOP trapping beam. We provide experimental details of the most important aspects, such as ensuring the spatial overlap of the two orthogonally-polarized components of the tweezer beam and the calibration of the DOP in the trapping region. 

\begin{figure}[ht]
    \centering
    \includegraphics[width=0.96\textwidth
    ]{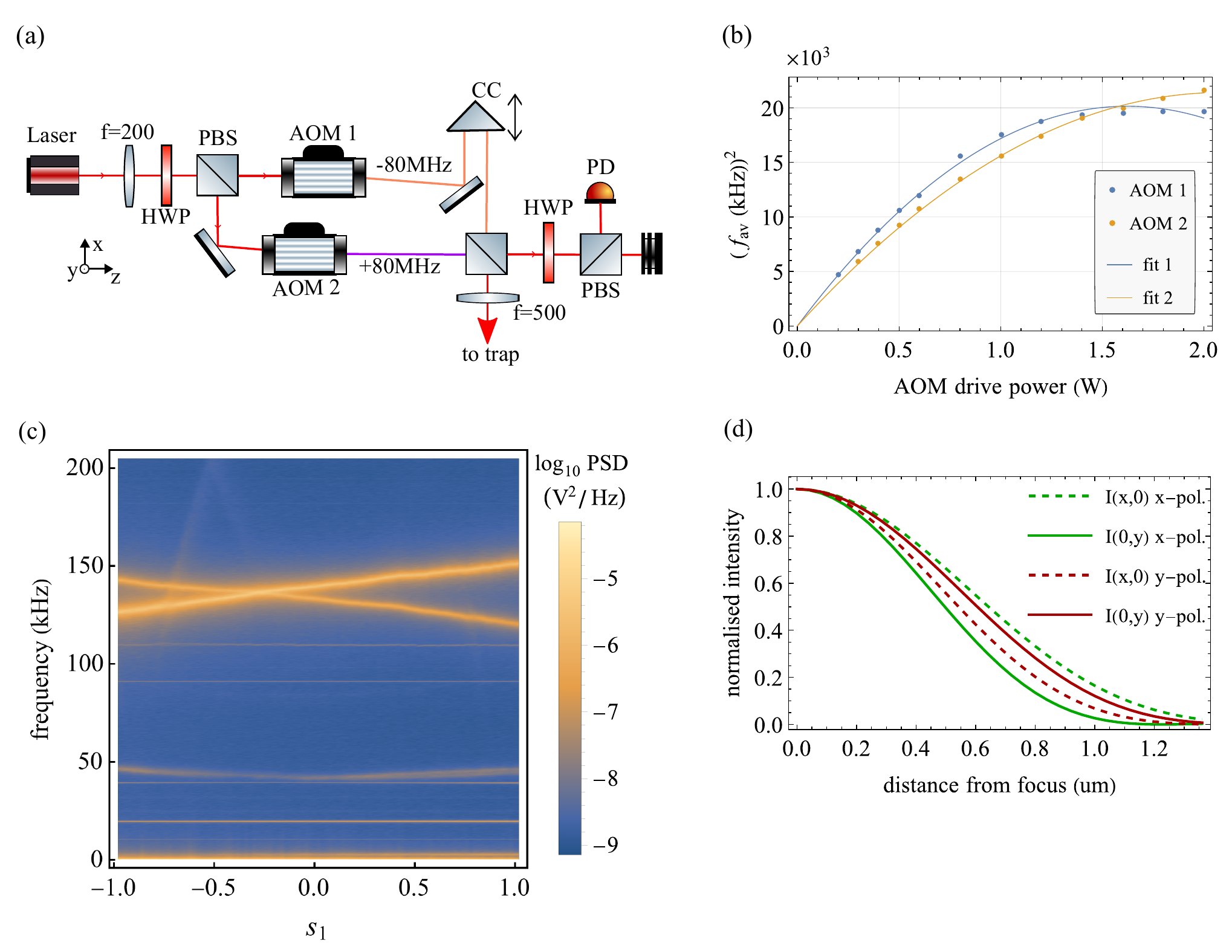}
    \caption{(a)~Experimental diagram of the beam preparation stage. The laser beam is split into two parts, which are frequency-shifted by $\pm80$~MHz (shown by violet and orange). The two parts are subsequently recombined and sent to the trap. Abbreviations:  half-wave plate (HWP), polarizing beam splitter (PBS), photo-diode (PD), corner cube (CC). (b)~The average COM transverse frequency squared $f_{\rm av}^2$ as a function of AOM driving power for both AOMs. The data points are shown together with quadratic fits (solid lines). 
    (c)~The power spectral density (PSD) of COM motion as a function of DOP of the trapping beam. This subfigure consists of 100 PSDs, where the first one ($s_1=-1$) corresponds to $y$-polarized trapping light and the last one ($s_1=1$) to $x$-polarized trapping light. The PSD is measured by the quadrant photo detector (see Fig.~\ref{fig:Setup}(a), both QPD channels are summed here). COM transverse frequencies $f_x$ and $f_y$ are visible in the 120 to 150~kHz region.
    (d)~Simulated intensity profiles along $x$ and $y$ directions for a strongly focused elliptical beam. Note that the cross-section of the focal spot is more circular for $y$-polarised tweezer beam.}
    \label{fig:COMcal}
\end{figure}

The tweezer beam generation system is depicted in Fig.~ \ref{fig:COMcal}(a). Before entering the vacuum chamber, the trapping beam, propagating along the $z$ axis, is split into two components with equal power but orthogonal linear polarizations. Each of these constituent beams is then frequency shifted by an AOM ($y$-polarized beam is shifted by $-80$~MHz by AOM~1, and $x$-polarized beam is shifted by $+80$~MHz by AOM ~2). Subsequently, the beams are spatially overlapped and recombined on a polarizing beam-splitter (PBS). Both beams are first focused into their respective AOMs using one lens before the split (f~$=200$~mm), and then collimated using another lens (f~$=500$~mm) after the 
recombination. The spatial mode overlap of the two beams is optimised by maximising the visibility of interference at the difference frequency $\Delta\omega=160$~MHz, recorded on the auxiliary detector (PD in Fig. \ref{fig:COMcal}(a)). The beams are overlapped by means of two steering mirrors, and the beam sizes are matched using a corner-cube (CC) mounted on a translation stage. The CC allows us to adjust the path length difference between the $x$- and $y$-polarized components between focusing (f~$=200$~mm) and collimating (f~$=500$~mm) lenses. The maximum achieved interference visibility is $85\%$.

The aforementioned system allows us to control the relative contributions of $x$- and $y$-polarized light to the trapping beam via controlling the  driving power of the AOMs. For example, if AOM~1 is driven at maximum power, and AOM~2 receives no drive, the tweezer is $y$ polarized, whereas when both AOM~1 and AOM~2 are driven so that they provide the same diffraction efficiency, the tweezer is composed of $x$ and $y$ polarizations in equal measure.

In the remainder of this section we will describe the calibration of the contributions of $x$- and $y$-polarized light to the intensity in the trapping volume, which directly determine the DOP in the trap. In order to avoid systematic errors introduced by spatial mode mismatch of the $x$- and $y$-polarized beams, we perform this calibration using the COM motion of the trapped particle.

The relationship between the focal intensity and the AOM driving power is determined for each beam separately with the help of the transverse COM frequencies describing the particle motion in the focal plane along $x$ and $y$ (denoted as $f_x$  and $f_y$ respectively). Let us denote the average transverse COM frequency as $f_{\rm av}=\frac{1}{2}(f_x +f_y)$. We use the average transverse COM frequency squared $f_{\rm av}^2$ as a proxy for the focal intensity measurement. 

The measured dependence of $f_{\rm av}^2$ on the driving power of each AOM (together with a quadratic fit) is depicted in Fig.~\ref{fig:COMcal}(b). We verify the calibration by performing the scan of the degree of polarization shown in Fig.~\ref{fig:COMcal}(c). As desired, we find that $f_{\rm av}$ remains constant when $s_1$ is varied. Note that the average transverse COM frequency $f_{\rm av}$ does not decrease for $s_1=0$ (when $50\%$ of focal intensity is $x$-polarized and $ 50\%$ is $y$-polarized), even though the calibration is performed for each polarization component separately. This shows that the volumes of the traps generated by the two beams overlap well and their trap depths add together to form the final trapping potential.

The DOP calibration procedure is potentially affected by the fact that in our experiment the trapping beam has a slightly elliptical shape introduced by the AOMs. For a tweezer formed by a strongly focused circular Gaussian input beam, $f_x$ and $f_y$ are not the same~\cite{novotny_hecht_2012}. We quantify this effect using $\epsilon=(f_x-f_y)/(f_x+f_y)$. Clearly, in the case of circular Gaussian beam, the value of $|\epsilon|$ should not depend on whether the tweezer is $x$- or $y$-polarized. Therefore, different values of $|\epsilon|$ for $x$ and $y$ tweezer polarization (see Fig.~\ref{fig:COMcal}(c) at points $s_1=\pm1$) suggest that the beam cross section is elliptical. 

In order to quantify the effect of the trapping beam ellipticity on $f_{\rm av}^2$, we have simulated the focal field produced by the trapping lens~\cite{novotny_hecht_2012}. The beam waists along $x$ and $y$ before focusing are used as free parameters. The obtained focal intensity cross sections along $x$ and $y$ for both $x$ and $y$ polarized tweezer light are shown in Fig.~\ref{fig:COMcal}(d) and reproduce well the observed values of $\epsilon$ for an $x$- and $y$-polarized tweezer ($\epsilon=0.057$  and $\epsilon=-0.116$ respectively). We find that the values of the average transverse COM frequency $f_{\rm av}$ for both polarizations differs by less than $1\%$. Hence, we conclude that $f_{\rm av}^2$ offers a good focal intensity estimate in our experiment.

\section{S2. Librational potentials}
 \label{s:potential}
In this section we derive the potential governing the libration of an anisotropic particle. The potential is induced by the electric field of the trapping beam which is propagating along $z$ and is composed of two frequency shifted  components ($x$- and $y$-polarized) of different amplitudes:
\begin{equation}
    \mathbf{E}=(E_x e^{i \omega_x t},E_y e^{i \omega_y t},0)^T.
    \label{Eq:Efield}
\end{equation}
The $x$ and $y$ field components acquire a phase difference that grows linearly in time in proportion to the frequency $\Delta\omega=\omega_x -\omega_y$. Since the initial phase difference is not important to the dynamics, we assume that the amplitudes $E_x$ and $E_y$ are real.

We assume that both the moment of inertia and the polarizability tensor can be simultaneously diagonalized in the principal axes reference frame of the particle. We refer to this frame of reference as  ``particle frame'' and represent the principal axes as $(e_1, e_2, e_3)$. We denote $\boldsymbol{\alpha}=\text{diag} (\alpha_1,\alpha_2,\alpha_3)$ and $\boldsymbol{I}= \text{diag} (I_1,I_2,I_3)$ as the static polarizability and inertia tensors of the object in the intrinsic body frame, respectively, assuming $\alpha_1 \leq \alpha_2 \leq \alpha_3$ and $I_1 \geq I_2 \geq I_3$. We refer to $e_3$ (the principal axis with the largest polarizability) as the ``long axis'' of the object.

In order to describe the orientation of the particle frame with respect to the laboratory frame ($x,y,z$), we use the intrinsic $x$-convention of Euler angles denoted as $\phi$, $\theta$ and $\psi$ (see  \cite{WolframEuler} and \S  35 in~\cite{Landau1976Mechanics}). The Euler angles $\phi$ and $\theta$ describe the orientation of the long axis of the rotor (the angle measured in the experiment is $\phi$, which corresponds to the orientation of the long axis in the $xy$ plane). In order to transform a vector from laboratory to particle frame we first rotate it by the angle $\phi$ around $z$, then by $\theta$ around $e_1$ and finally by $\psi$ around $e_3$. The transformation matrix corresponding to these three rotation operations   reads~\cite{WolframEuler} 
\begin{equation}
    \bf{R}=\left(
\arraycolsep=10pt
\begin{array}{ccc} 
 \cos \psi \cos \phi -\cos \theta  \sin \psi  \sin \phi  & \cos \theta  \sin \psi  \cos \phi +\cos \psi  \sin \phi  & \sin \theta  \sin \psi  \\
 -\cos \theta  \cos \psi  \sin \phi -\sin \psi  \cos \phi  & \cos \theta  \cos \psi  \cos \phi -\sin \psi  \sin \phi  & \sin \theta  \cos \psi  \\
 \sin \theta  \sin \phi  & -\sin \theta  \cos \phi  & \cos \theta \\
\end{array}
\right) .
\end{equation}

The dipole moment induced in a trapped anisotropic particle, expressed in the laboratory frame of reference, yields:  
\begin{equation}
    \mathbf{p}=\boldsymbol{R^{-1} \alpha R} \mathbf{E}, 
\end{equation}
where 
$\vec{E}$ is also expressed in the laboratory reference frame.  
Since in general $\vec{p}$ and $\vec{E}$ are not parallel, the potential energy $ U_{\rm tot}$ associated with the orientation of the particle  (after averaging over optical frequencies) yields: 
\begin{equation}
    U_{\rm tot}=-\frac{1}{4}\text{Re} \left(\mathbf{p}\cdot \mathbf{E}^*\right).
\end{equation}
If we consider the electric field as in Eq.~\eqref{Eq:Efield}, the potential $ U_{\rm tot}$ can be written as a sum of two terms
\begin{equation}
   U_{\rm tot}=U_0 + U_{1} \cos{(\Delta\omega t)} ,
\end{equation}
with
\begin{subequations}
\begin{align}
   U_{0} =& \frac{S_0}{8} \bigl[-(\alpha _1+\alpha _2) +\sin ^2{\theta}\; (1-s_1 \cos{2 \phi}) \left(\alpha _1 \sin ^2{\psi}+\alpha _2 \cos ^2{\psi}-\alpha _3\right)\nonumber\\
   &+ s_1 \left(\alpha _1-\alpha _2\right) (\cos{\theta}\sin {2 \psi} \sin{2 \phi}-\cos{2 \psi} \cos{2 \phi})\bigr] , \label{eq:U0}\\
    U_1 =& -\frac{S_0 \sqrt{1-s_1^2}}{8} \;\bigl[\sin ^2{\theta} \sin{2 \phi} \;(\alpha _1 \sin ^2{\psi}+\alpha _2 \cos ^2{\psi}-\alpha _3)\nonumber\\
    &+(\alpha _1-\alpha _2) (\cos{\theta} \sin{2 \psi} \cos{2 \phi} +\cos{2 \psi} \sin{2 \phi})\bigr],
\end{align}
\end{subequations}
where $S_0$ and $s_1=S_1/S_0$ are defined in Eqs.~\eqref{eq:S0} and ~\eqref{eq:S1}.
The oscillating term $U_{\rm 1}\cos(\Delta\omega t)$ describes the interaction of a dipole moment induced by the field oscillating at $\omega_x$ with the field oscillating at $\omega_y$ (and vice versa).

The fast oscillating term $U_{1} \cos(\Delta\omega t)$  will give rise to a small amplitude micromotion at $\Delta\omega$. This ``fast''  micromotion will in turn have an effect on the ``slow'' librational dynamics, which can be described as an additional, constant-in-time effective  potential term  $U_1'$ ~\cite{Landau1976Mechanics, RahavPRA2003}, yielding
\begin{equation}
   U_{\rm 1}'=\frac{1}{2 (\Delta\omega)^2}\sum_{i,k} \mathcal{A}_{ik}^{-1}\partial_i U_1 \partial_k U_1 \;,
\end{equation}
where the indices $i$ and $k$ run through $\theta, \phi$ and $\psi$ and $\mathcal{A}_{ik}$ are matrix elements of the quadratic form describing the kinetic energy $T$, such that
\begin{equation}
    T=\frac{1}{2}(\dot\theta, \dot\phi,\dot\psi)\mathbfcal{A} (\dot\theta, \dot\phi,\dot\psi)^{\rm T}.
\end{equation}
The matrix $\mathbfcal{A}$ depends on the Euler angles and the inertial moment according to 
\begin{equation}
   \mathbfcal{A}= \left(
\arraycolsep=10pt
\begin{array}{ccc}
 I_1 \cos ^2{\psi}+I_2 \sin ^2{\psi} & (I_1-I_2)\sin{\theta} \sin {\psi} \cos{\psi} & 0 \\
(I_1-I_2)  \sin{\theta}\sin {\psi} \cos{\psi} &  (I_1 \sin ^2{\psi}+I_2 \cos ^2{\psi})\sin ^2{\theta}+I_3 \cos ^2{\theta} & I_3 \cos{\theta} \\
 0 & I_3 \cos{\theta} & I_3 \\
\end{array}
\right).
\end{equation}

As an example, let us find the explicit expressions for $U_0$ and $U_1'$ in the case of  a rotor with cylindrical symmetry, such as a dumbbell ($\alpha_1=\alpha_2$, $I_1=I_2$): 
\begin{subequations}
\begin{align}
  U_{0} &=\frac{A_1^2}{4}\;I_1 (s_1 \cos{2 \phi}-1)\; \sin ^2{\theta} \;,\label{eq:u0dumb} \\
U_{1}'&=\frac{A_1^4 I_1(1-s_1^2) \sin^2{\theta}}{8 (\Delta\omega)^2 } \; \frac{\bigl[ I_1\sin ^2{\theta} \cos^2{2\phi}+(I_1\sin ^2{\theta}+I_3 \cos ^2{\theta}) \cos ^2{\theta} \sin ^2{2 \phi} \bigr]}{ (I_1 \sin ^2{\theta}+I_3\cos ^2{\theta })}\;,\label{eq:u1prim}
\end{align}
\end{subequations}
where the parameter $A_1=\sqrt{\frac{(\alpha _3-\alpha _1) S_0}{2 I_1}}$ is equal to the libration frequency in a linearly polarized trap. 
Equations~\eqref{eq:u0dumb} and ~\eqref{eq:u1prim} indicate that the residual potential $U_1'$ is shallower than $U_0$ by approximately a factor of $(A_1/\Delta\omega)^2$. Note that for our experimental parameters, we have $(A_1/\Delta\omega)^2\approx 10^{-5}$.

Let us now consider how both potential terms affect the dynamics of the orientation of the dumbbell in the $xy$ plane described by the angle $\phi$. We can expect the contribution from $U_1'$ to be negligible, except when  $s_1\approx0$ (the trap is unpolarized) and $U_0$ is independent of $\phi$. Therefore, in this case, $U_1'$ is the dominant potential term governing the dynamics of $\phi$. The minima of $U_1'$ occur at $\phi=n \pi/4$ with $n\in\{1,2,3,4\}$, which means that our symmetric rotor will become diagonally oriented in the $xy$ plane.

However, in our experiment we have $U_1'\ll k_\mathrm{B} T$, where $k_\mathrm{B}$ is the Boltzmann constant and $T$ is the temperature. Therefore, even when the tweezer is unpolarized, the effect of the residual potential $U_1'$  is completely obscured by the interaction with the environment (background gas).

\section{S3. Torques and libration frequencies for asymmetric rotor}
 \label{s:equationsofmotion}
In this section we use the potential derived in the previous section to calculate restoring torques acting on a rotor trapped in a beam with an arbitrary DOP. We derive the libration frequencies as functions of DOP.

We denote particle-frame torque components as $K_i$ and  $k_i=\frac{K_i}{I_i}$. The full expressions for torques due to the time-independent potential $U_{0}$ [see Eq.~\eqref{eq:U0}] read
\begin{subequations}
\begin{align}
   k_1 &= k_{\rm N} \cos\psi +\frac{(k_z-k_3\cos\theta)\sin\psi}{\sin\theta} = \frac{A_1^2}{2} \sin \theta  \left[(1-s_1 \cos 2 \phi ) \cos \theta  \cos \psi  + s_1 \sin \psi  \sin 2\phi  \right] \;, \label{eq:torqueFull1}\\
    k_2 &=-k_{\rm N} \sin\psi +\frac{(k_z-k_3\cos\theta)\cos\psi}{\sin\theta} = -\frac{A_2^2 }{2} \sin \theta  [(1-s_1 \cos 2 \phi )\cos \theta  \sin \psi  -s_1  \cos \psi  \sin 2 \phi ]\;,\\
    k_3 & =-\frac{A_3^2 }{2}  \left[(1-s_1 \cos 2 \phi )\sin^2{\theta} \sin{\psi} \cos\psi +s_1 (\cos\theta \cos 2 \psi \sin 2 \phi +\sin 2 \psi \cos 2 \phi )\right],
    \label{eq:torqueFull}
\end{align}
\end{subequations}
where $k_3=-\partial U_0/ \partial \psi$, $k_{\rm N}=-\partial U_0/ \partial \theta$, $k_z=-\partial U_0/ \partial \phi$. Additionally we have $A_i=[(|\alpha _j-\alpha _k| S_0)/2 I_i]^{1/2}$ where  $\{i,j,k\}$ are permutations of $\{1,2,3\}$. 

To gain an intuitive understanding, let us analyze in detail the case $s_1=1$, i.e., the tweezer is linearly polarized along $x$. The object is trapped in a potential minimum (along $x$), for which $\theta=\phi=\frac{\pi}{2}$, and $\psi$ is unrestrained (free). This means that the rotor aligns itself with its long axis to the polarization axis, while it can freely spin around it. 
Let us now move beyond purely linear polarization. As soon as $s_1<1$, a potential minimum appears at $\psi=\frac{\pi}{2}$ and the angle $\psi$ also becomes trapped. Colloquially speaking, the rotor can now minimize its potential energy by ``lying flat'' in the polarization plane.  
Having discussed the orientation with minimal potential energy, let us consider deviations from that orientation and the associated dynamics, which corresponds to librational motion for small angles.
To this end, we introduce the libration angles describing the motion around the equilibrium position as $\Theta=\theta-\frac{\pi}{2},  \Phi=\phi-\frac{\pi}{2}$, and $\Psi=\psi-\frac{\pi}{2}$. 
Without loss of generality, it is convenient to assume a tweezer field that is predominantly $x$-polarized, such that the resulting torque components can be expressed in the laboratory frame. To first order in the Euler angles, these restoring torques acting on the angles $\Theta$, $\Phi$, and $\Psi$ read
\begin{subequations}
\begin{align}
    k_z&\approx - s_1 A_1^2 \Phi \label{eq:torqueR1},\\
    k_y &\approx -\frac{s_1+1}{2}A_2^2 \Theta \label{eq:torqueR2},\\
    k_x &\approx-\frac{1-s_1}{2} A_3^2 \Psi. \label{eq:torqueS}
\end{align}
\end{subequations} 
The libration frequencies associated with these restoring torques read
\begin{subequations}\label{eq:librationFreq}
\begin{align}
   \Omega_1&= \sqrt{\mathcal{P}}A_1 \label{eq:freqS1},\\
   \Omega_2&= \sqrt{\frac{\mathcal{P}+1}{2}}A_2 \label{eq:freqS2},\\
    \Omega_3&=\sqrt{\frac{1-\mathcal{P}}{2}} A_3.\label{eq:freqS3}
\end{align}
\end{subequations}
In the above expressions we have replaced $s_1$ with $\mathcal{P}$ so that Eqs.~\eqref{eq:freqS1}--\eqref{eq:freqS3} are also valid around $s_1=-1$. In the case of $s_1\approx-1$ the tweezer is almost $y$-polarized, and the potential minimum occurs for a different particle orientation. Therefore, the angles $\theta$, $\phi$ and $\psi$ librate around different equilibrium positions (denoting the potential minimum) for different polarization states of the tweezer field. However, the libration frequencies are, to linear order, always given by Eq.~\eqref{eq:librationFreq}.

\section{S4. Equations of motion for a symmetric rotor}
 \label{s:equationsofmotionDumb}

In this section, we analyze the motion of a symmetric rotor ($I_1= I_2$,  $\alpha_1= \alpha_2$) under restoring torques given by Eqs.~\eqref{eq:torqueFull1}--\eqref{eq:torqueFull} and an additional small spinning torque. Since the parameter $A_3$ vanishes for the symmetric rotor, Eq.~\eqref{eq:torqueFull} implies that there is no restoring torque pointing along the object's long axis. Therefore, for symmetric rotors the angle $\psi$ is free for any value of $s_1$. 

Let us now explore a scenario in which the symmetric rotor is performing a small amplitude libration around the $x$ axis ($s_1\approx1$) and an additional constant torque is present. We assume that the additional torque is much smaller than the restoring torques along $y$ and $z$ [see Eqs.~\eqref{eq:torqueR1} and~\eqref{eq:torqueR2}]. Therefore, only the additional torque component pointing along $x$ will have an effect on the dynamics, causing the rotor to spin around its long axis (which points long $x$).  We denote the additional torque component along $x$ as $k_{{\rm ext}}$.  
Let us write the rotational equations of motion~\cite{Landau1976Mechanics} (Euler's equations), including both the restoring torques and $k_{{\rm ext}}$ for a symmetric rotor performing small amplitude libration around the $x$ axis :
\begin{subequations}
\begin{align}
   \ddot \Phi &= -\mu \dot \Theta \dot \Psi- \Omega_1^2 \Phi \label{eq:motion1},\\
   \ddot \Theta &= \mu \dot \Psi \dot\Phi - \Omega_2^2 \Theta \label{eq:motion2} ,\\
   \ddot \Psi&= k_{{\rm ext}},
\end{align}
\end{subequations}
where $\mu=\frac{I_1-I_3}{I_1}$. Provided that $k_{{\rm ext}}$ is larger than fluctuating (thermal) torques, the dumbbell will start to spin consistently around its long axis. The spinning rate will increase until it reaches the friction-dependent stationary value denoted as $\dot\Psi_0$.  Note that Eqs.~\eqref{eq:motion1} and~\eqref{eq:motion2} include coupling between angular degrees of freedom, which is an inherent feature of rotational mechanics. Fast spinning, i.e., large $\dot\Psi$  causes strong coupling between the $\Phi$ and $\Theta$ libration modes which become hybrid precession modes. Similarly to Ref.~\cite{SebersonPRA2019} we use the following simple model
\begin{subequations}
\begin{align}
   \ddot \Phi &= -\omega_{\rm s} \dot \Theta - \Omega_1^2 \Phi \label{eq:spinmodel1},\\
   \ddot \Theta &= \omega_{\rm s}  \dot\Phi  -\Omega_2^2 \Theta,
   \label{eq:spinmodel2}
\end{align}
\end{subequations}
where $\omega_{\rm s}=\mu \dot\Psi_0$. We will henceforth refer to $\omega_{\rm s}$ as the rotational coupling rate.  The eigenfunctions of Eqs.~\eqref{eq:spinmodel1}-\eqref{eq:spinmodel2} (precession modes) oscillate at eigenfrequencies $\Omega_{\rm A}$ and $\Omega_{\rm B}$, which read
\begin{subequations}
\begin{align}
   \Omega_{\rm A}&= \frac{\sqrt{\Omega_1^2+\Omega_2^2+\omega_{\rm s}^2-\sqrt{\left(\Omega_1^2+\Omega_2^2+\omega_{\rm s} ^2\right)^2-4 \Omega_1^2 \Omega_2^2}}}{\sqrt{2}} \label{eq:omegaA}, \\
   \Omega_{\rm B}&= \frac{\sqrt{\Omega_1^2+\Omega_2^2+\omega_{\rm s}^2+\sqrt{\left(\Omega_1^2+\Omega_2^2+\omega_{\rm s} ^2\right)^2-4 \Omega_1^2 \Omega_2^2}}}{\sqrt{2}}.
   \label{eq:omegaB}
\end{align}
\end{subequations}
We notice that the precession frequencies $\Omega_{\rm A}$ and $\Omega_{\rm B}$ satisfy the following relation:
\begin{align}
(\Omega_{\rm A}-\Omega_{\rm B})^2= \omega_{\rm s}^2+(\Omega_1-\Omega_2)^2.
\label{eq:ExcSpin}
\end{align}
This relation illustrates the signature of the spinning along the long axis, which can be described as an additional splitting between the eigenfrequencies visible in the libration spectrum. We have used Eq.~\eqref{eq:ExcSpin} to reconstruct the value of the rotational coupling $\omega_{\rm s}$ (as function of $s_1$) from the measured precession mode splitting $(\Omega_{\rm A}-\Omega_{\rm B})$ and the libration mode splitting $(\Omega_{1}-\Omega_{2})$ predicted from Eqs.~\eqref{eq:freqS1} and~\eqref{eq:freqS2}. The value of $A_1$ used in this procedure was measured at $\mathcal{P}=0$. The obtained rotational coupling rate is shown in Fig.~\ref{fig:ramps}(b) as a dashed black line.

Note that $\omega_{\rm s}$ depends only on the difference between observed peak frequencies in the libration spectrum. We have calculated $\omega_{\rm s}$ from the data in Fig.~\ref{fig:ramps}(b) and used it to find the precession mode frequencies from Eqs.~\eqref{eq:omegaA} and~\eqref{eq:omegaB}. We find that the resulting precession frequencies (as functions of $s_1$) fit the behavior of the modes observed in the libration spectrum well, which confirms the spinning hypothesis. 

Lastly, we have only discussed the effect of the additional small torque on a symmetric rotor (dumbbell). We note that non-symmetric rotors (clusters), when exposed to the same additional torque, will not spin. This is due to the fact that for $\mathcal{P}<1$ all degrees of freedom describing cluster orientation are trapped in a potential minimum, such that they librate and do not spin, unless the additional torque exceeds the restoring torque.
\end{document}